\documentclass[aps,prl,twocolumn,superscriptaddress,showpacs]{revtex4}
\usepackage{graphicx}
\usepackage{mathrsfs}

\begin{document}



\title{Current-induced cleaning of graphene}


\author{J.~Moser}
\email{moser.joel@gmail.com}
\affiliation{CIN2 and CNM-CSIC, Campus Universitat Autonoma de
Barcelona, E-08193 Bellaterra, Spain\\}

\author{A.~Barreiro}
\affiliation{CIN2 and CNM-CSIC,
Campus Universitat Autonoma de Barcelona, E-08193 Bellaterra,
Spain\\}

\author{A.~Bachtold}
\email{adrian.bachtold@cnm.es}
\affiliation{CIN2 and CNM-CSIC, Campus
Universitat Autonoma de Barcelona, E-08193 Bellaterra, Spain\\}


\begin{abstract}

A simple yet highly reproducible method to suppress contamination of
graphene at low temperature inside the cryostat is presented. The
method consists of applying a current of several mA through the
graphene device, which is here typically a few $\mu$m wide. This
ultra-high current density is shown to remove contamination adsorbed
on the surface. This method is well suited for quantum electron
transport studies of undoped graphene devices, and its utility is
demonstrated here by measuring the anomalous quantum Hall effect.

\end{abstract}


\maketitle

The discovery of graphene, a one-atom thick layer of graphite, has
generated considerable interest by opening avenues in condensed
matter physics \cite{review}. Graphene is an appealing system for
basic research due to its unusual electronic properties; namely,
charge carriers behave like chiral massless Dirac particles
\cite{QHE1,QHE2,QHE3,QHE4}. The material also holds promise for
technological applications, such as in nanoelectronics
\cite{deHeer}; in addition, graphene has been shown to be an
outstanding gas sensor able to detect minute amounts of molecular
dopants \cite{sensor1,sensor2}. Yet, what makes graphene an enticing
material is also the source of great technological difficulties:
being in essence a surface, graphene proves to be extremely
sensitive to contamination.

Recently, different approaches have been put forward to address this
problem. Annealing at several hundreds Celsius in ultra-high vacuum
or in Ar/H$_{2}$ environment has been shown to remove contamination
\cite{fuhrer_nanoletters,PNAS_Kim}. Unfortunately, however, such
annealing processes are difficult to implement in a cryostat for
low-temperature measurements, and transferring samples from the
annealing chamber to the cryostat is not convenient, since exposure
to air reintroduces contamination.

\begin{figure}[t]
\includegraphics{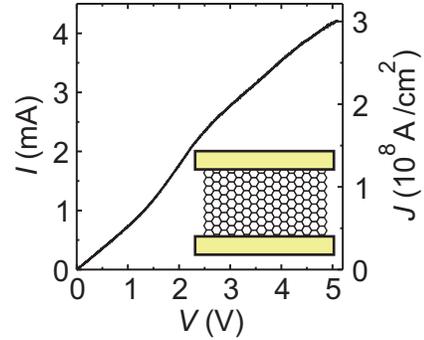}
\caption{(Color online) current $I$ (left axis) and current density
$J$ (right axis) as a function of source-drain bias $V$ applied
across the graphene sample in helium atmosphere at $T=76\,$K.
Graphene width $w=4\,\mu$m and inter-electrode distance $L=1\,\mu$m.
Inset: schematic of the device.}
\end{figure}

In this Letter, a simple yet highly reproducible method to suppress
contamination at low temperature inside the cryostat is presented.
The method is based on the application of a large current through
the graphene device generating several tens of mW dissipation over a
few $\mu$m$^2$ large surface. Remarkably, graphene can sustain such
extreme conditions while adsorbed contamination gets removed. We
employ atomic force microscopy (AFM) to show how adsorbates, as well
as intentionally deposited CdSe particles, melt away. These cleaned
graphene devices allow us to measure the anomalous quantum Hall
effect.

We start by briefly describing our fabrication technique. Graphene
flakes are deposited by mechanical exfoliation of Kish graphite
(Toshiba Ceramics) on degenerately doped silicon substrates coated
with $270\,$nm of thermal silicon oxide. We use standard wafer
protection tape as it leaves little adhesive residue on substrates.
Cr/Au electrodes are fabricated on top of samples by electron beam
lithography followed by lift-off in acetone and dicholoroethane.
Two-point {\sl dc} transport measurements are conducted both at room
temperature in air for AFM imaging and in a He$^4$ cryostat.

Graphene is able to carry very high electrical currents without
sustaining damage. Applying a source-drain bias $V$ of a few Volts
across the sample induces a large current flow $I$ of a few mA, as
shown in Fig.~1. Taking the sample width of $4\,\mu$m and a
thickness of $0.35\,$nm, this translates into an extremely large
current density $J$ of a few $10^8\,$A/cm$^2$. For comparison, $J$
is only a few times larger in carbon nanotubes
\cite{dekker,J_in_nanotubes}, and for both materials $J$ is several
orders of magnitude larger than in present-day interconnects.

\begin{figure}[t]
\includegraphics{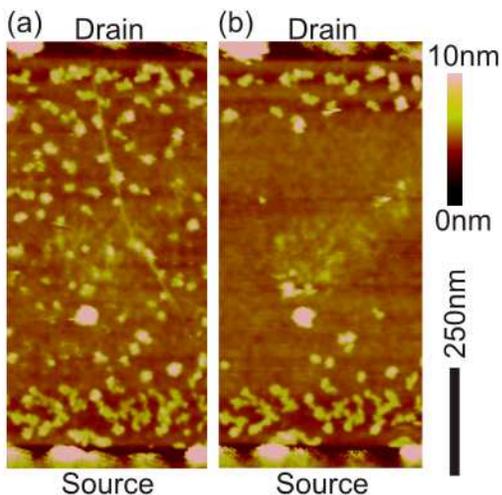}
\caption{(Color online) CdSe nanoparticles are deposited on a
graphene sample, and imaged by AFM before (Fig.~2a) and after
(Fig.~2b) applying a large electrical current between source and
drain. Both Figs. are topographic scans. Fig.~2b indicates that most
nanoparticles have vanished from the area located halfway between
source and drain. Particles are not observed to cluster up nor to
migrate towards the edges over the entire width of the electrodes
(not shown).}
\end{figure}

The effects of a large electrical current passing through a
mesoscopic, conducting device include electromigration
\cite{EM_Hummel,EM_mceuen} and Joule heating. We expect that the
large power $P\simeq 20\,$mW dissipated over a small area of a few
$\mu$m$^2$ significantly increases the sample temperature. In order
to show this, we monitor the four-point resistance $R$ of a gold
strip ($R\simeq 60\,\Omega$ at room $T$) fabricated on the substrate
$30\,\mu$m away from the graphene sample. Having initially
calibrated $R$ against a commercial thermometer in thermal
equilibrium, we estimate that $T$ rises by $\sim 25\,$K at the gold
strip when a power of $34\,$mW is dissipated in the graphene sample
at $4.2\,$K. However an estimate of the graphene temperature is
difficult to obtain, since most of the heat is evacuated through the
Au electrodes that are thermally anchored to the liquid helium bath
(see below). Further insight is gained by depositing CdSe
nanoparticles with a mean diameter of $5\,$nm on a graphene sample
and applying a large current. Fig.~2 presents topographic AFM scans
of the sample before (Fig.~2a) and after (Fig.~2b) the current flow.
Fig.~2b shows that most nanoparticles have vanished. Owing to high
temperatures generated by the large Joule heating, nanoparticles
might have undergone various phase transitions: they might have
melted and formed a thin film over the sample; they might also have
evaporated or sublimated. In addition, each of these mechanisms may
be assisted by electromigration \cite{zettl}. A rough estimate for
the temperature reached is given by the melting temperature of
$5\,$nm diameter CdSe particles, which is expected to be $\sim
600\,^\circ$C \cite{footnote}. Future work should help identify the
mechanism by which nanoparticles disappear and allow for a better
estimate of the graphene temperature. Eventually, note that the
presence of remaining particles close to the electrodes suggests
that these act as heat sinks (Fig.~2b).

\begin{figure}[t]
\includegraphics{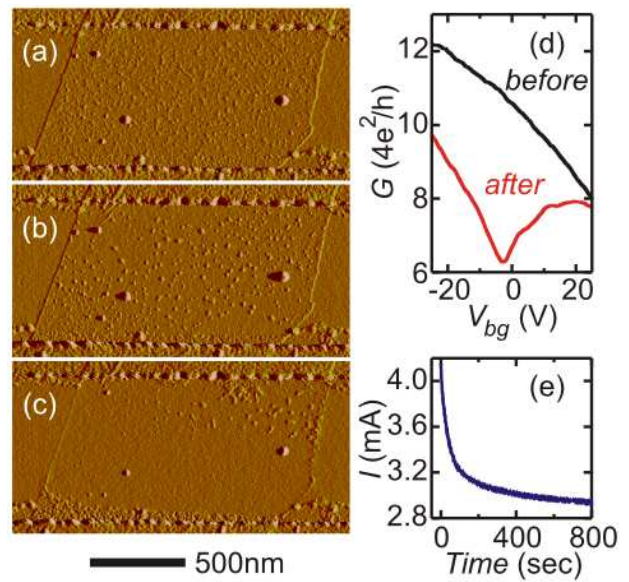}
\caption{(Color online) (a)-(c) AFM signal amplitude scans captured
at various stages of the current-induced cleaning process in air.
(d) Two-point conductance $G$ as a function of back gate bias
$V_{bg}$ before and after the application of a large current, in
helium atmosphere and at $T=76\,$K. (e) Time dependence of current
$I$ at fixed bias $V=5\,$V.}
\end{figure}

In the following we demonstrate how this effect can be harnessed to
remove contaminants from graphene samples. AFM imaging reveals the
influence of the large electrical current on the morphology of the
sample surface. Figs.~3a-b display AFM scans for the signal
amplitude at various stages of the cleaning process. Typically,
freshly deposited graphene sheets have an average roughness
$R_{a}\simeq 0.1\,$nm. Right after fabrication (Fig.~3a)
$R_{a}\simeq 0.4\,$nm is measured, suggesting the presence of
contamination on top of graphene. Fig.~3b captures an intermediate
state obtained by ramping $V$ to $3\,$V. Contaminants migrate and
cluster up. As $V$ is further increased, the sample area halfway
between source and drain becomes almost as smooth as the substrate
(Fig.~3c, for $V=4.6\,$V) and the roughness is down to $R_{a}\simeq
0.1\,$nm, as before fabrication. Fig.~3c also shows that some
contamination has accumulated at the corner of the device (such a
migration is not observed with CdSe particles).

This contamination removal has drastic effects on the sample doping.
Electric field effect can be used to dope graphene either with
electrons or holes, resulting in a minimum conductivity at the
charge neutrality point. The black curve in Fig.~3d represents the
low-bias conductance $G$ as a function of a back gate bias $V_{bg}$
applied to the silicon substrate, for the same sample as in Fig.~1
but prior to any large current flow. Measurements are performed
under helium atmosphere at $T=76\,$K in a cryostat. $G$ decreases
continuously as a function of $V_{bg}$, which indicates that the
conductance minimum is beyond the voltage range and that doping is
important. After high-bias treatment we observe a well-defined
conductance minimum $G_{min}$ close to zero $V_{bg}$ that denotes
the reduction of doping.

The strong shift of $G_{min}$ to zero $V_{bg}$, along with the
recovered smoothness of the sample, indicates that unavoidable
fabrication residues act as dopants that are removed upon passing a
large electrical current through the sample. These residues may
consist in part of non-dissolved PMMA \cite{fuhrer_nanoletters}.
Note that in air, this cleaning process tends to be counterbalanced
by the doping effect of other molecular adsorbates ({\sl e.g.},
oxygen), as a result of which $G_{min}$ is found to slowly drift in
time back to large, positive $V_{bg}$ values.

Special care has to be taken to preserve the integrity of the device
upon applying large source-drain biases; we proceed as follows.
Setting $V_{bg}=0$ we slowly raise $V$. We regularly hold $V$
constant to monitor the time-dependence of the current $I$. If $I$
stays constant, we keep increasing $V$. This is repeated until $I$
continuously decreases as a function of time and saturates (see
Fig.~3e for a fixed $V=5\,$V, sample of Fig.~1 and Fig.~3d.) The
continuous decay of $I$ in time at fixed $V$ and $V_{bg}=0$ signals
a smooth shift of $G_{min}$ to zero $V_{bg}$. Note that an
acceleration of the current decay usually precedes the rupture of
graphene.

\begin{figure}[t]
\includegraphics{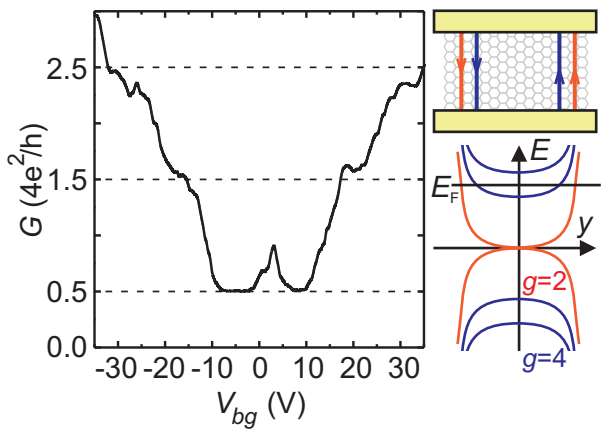}
\caption{(Color online) (Left) two-point conductance $G(V_{bg})$ at
$B=9\,$T and $T=4.2\,$K revealing the anomalous quantum Hall effect.
(Right) Edge channels in ideally clean, single layer graphene.
Electron and hole-like channels coexist close to the neutrality
point, yielding a channel degeneracy $g=2$. The $y$-axis is along
the sample width. Levels are bent at the edges due to hardwall
confinement.}
\end{figure}

Remarkably, this simple cleaning technique allows us to observe the
anomalous, half-integer quantum Hall effect (QHE) \cite{QHE1,QHE2}
that is the signature of a clean, single graphene layer. In
graphene, the $n^{\textrm{th}}$ Landau level (LL) energy reads
$E_n={\textrm{sgn}}(n)v_{\textrm{F}}\sqrt{2e\hbar |n|B}$, where $n$
is the electron or hole LL index and $v_{\textrm{F}}\simeq
10^6\,$m$\cdot$s$^{-1}$ is the Fermi velocity. Edge channels have a
degeneracy $g=4$ that accounts for both spin and sub-lattice
degeneracies. The presence of electron and hole edge channels at the
neutrality point (LL index $n=0$) reduces the number of
current-carrying modes by half, yielding $g=2$ (see schematic of
Fig.~4.) The two-point Landauer conductance reads $G=\sum g\cdot
e^{2}/h$, where the sum is carried over all the levels, and is
therefore quantized in $4e^{2}/h\cdot (n+1/2)$ as $V_{bg}$ is swept
\cite{delft}. Fig.~4 shows a clear quantization at $1/2\cdot
4e^{2}/h$ whereas the other plateaus are not as well defined (the
same is observed for 5 other samples). The robustness of the first
plateau is explained by the large energy gaps $|E_{\pm
1}-E_{0}|\simeq 1200\,$K at $B=9\,$T. Importantly, the plateau is
visible only after the current-induced cleaning process. This
indicates that our process enhances the mobility of the sample
and/or improves the homogeneity of the doping over the device area.

To conclude, we have demonstrated a simple, yet highly reproducible
technique to obtain clean graphene devices out of initially highly
contaminated samples. Our method takes advantage of the large
electrical current density that graphene can sustain to remove
contaminants from the sample surface. Compared to other methods, the
reported technique has the advantage to be operative at low $T$
inside a cryostat, without having to cycle the cryostat to room $T$.
In addition, we avoid contamination in air that would naturally
occur upon transferring the device from the annealing chamber to the
cryostat.

We thank P. Jarillo-Herrero for showing us the mechanical
exfoliation technique. This work is supported by an EURYI grant and
FP6-IST-021285-2.


\begin{references}

\bibitem{review}
For a review, see: A. Castro Neto, F. Guinea, and N. M. Peres,
Physics World, November 2006; A. K. Geim and K. S. Novoselov, Nature
materials {\bf 6}, 183 (2007); M. I. Katsnelson, Materials Today
{\bf 10}, 20 (2007).

\bibitem{QHE1}
K. S. Novoselov, A. K. Geim, S. V. Morozov, D. Jiang, M. I.
Katsnelson, I. V. Grigorieva, S. V. Dubonos, and A. A. Firsov,
Nature {\bf 438}, 197 (2005).

\bibitem{QHE2}
Y. Zhang, Y.-W. Tan, H. L. Stormer, and P. Kim, Nature {\bf 438},
201 (2005).

\bibitem{QHE3}
V. P. Gusynin and S. G. Sharapov, Phys. Rev. Lett. {\bf 95}, 146801
(2005).

\bibitem{QHE4}
N. M. R. Peres, F. Guinea, and A. H. Castro Neto, Phys. Rev. B {\bf
73}, 125411 (2006).

\bibitem{deHeer}
C. Berger, Z. Song, X. Li, X. Wu, N. Brown, C. Naud, D. Mayou, T.
Li, J. Hass, A. N. Marchenkov, E. H. Conrad, P. N. First, and W. A.
de Heer, Science {\bf 312}, 1191 (2006).

\bibitem{sensor1}
F. Schedin, K. S. Novoselov, S. V. Morozov, D. Jiang, E. H. Hill, P.
Blake, and A. K. Geim, cond-mat/0610809.

\bibitem{sensor2}
E. H. Hwang, S. Adam, S. Das Sarma, and A. K. Geim,
cond-mat/0610834v1

\bibitem{fuhrer_nanoletters}
M. Ishigami, J. H. Chen, W. G. Cullen, M. S. Fuhrer, and E. D.
Williams, Nano Lett. {\bf 7}, 1643 (2007).

\bibitem{PNAS_Kim}
E. Stolyarova, K. T. Rim, S. Ryu, J. Maultzsch, P. Kim, L. E. Brus,
T. F. Heinz, M. S. Hybertsen, and G. W. Flynn, PNAS {\bf 104}, 9209
(2007).

\bibitem{dekker}
Z. Yao, C. L. Kane, and C. Dekker, Phys. Rev. Lett. {\bf 84}, 2941
(2000).

\bibitem{J_in_nanotubes}
B. Bourlon, D. C. Glattli, B. Pla\c{c}ais, J. M. Berroir, C. Miko,
L. Forr\'{o}, and A. Bachtold, Phys. Rev. Lett. {\bf 92}, 026804
(2004).

\bibitem{EM_Hummel}
R. E. Hummel, International Materials Reviews {\bf 39}, 97 (1994).

\bibitem{EM_mceuen}
H. Park, A. K. L. Lim, J. Park, A. P. Alivisatos, and P. L. McEuen,
Appl. Phys. Lett. {\bf 75}, 301 (1999).

\bibitem{zettl}
B. C. Regan, S. Aloni, R. O. Ritchie, U. Dahmen, and A. Zettl,
Nature {\bf 428}, 924 (2004).

\bibitem{footnote}
The melting temperature $T_{m}$ of $4\,$nm diameter CdS particles
has been measured to be $1180\,$K [A. N. Goldstein, C. M. Echer, and
A. P. Alivisatos, Science {\bf 256}, 1425 (1992).] $T_{m}$ of CdSe
particles can be roughly estimated using $T_{m}/T_{m}^{bulk}=1-C/d$,
where $d$ is the particle diameter and $T_{m}^{bulk}$ is the bulk
material melting temperature ($1510\,$K for CdSe and $2020\,$K for
CdS) [Z. Zhang, M. Zhao, and Q. Jiang, Semicond. Sci. Technol. {\bf
16}, L33 (2001).] $C$ is material-dependent, yet here we take the
same value for CdSe and CdS.

\bibitem{delft}
H. B. Heersche, P. Jarillo-Herrero, J. B. Oostinga, L. M. K.
Vandersypen, and A. F. Morpurgo, Nature {\bf 446}, 56 (2007).

\end{references}
\end{document}